\documentclass[mathleft
]{an}
\usepackage{graphicx}
\usepackage{times}
\usepackage{amsmath}

\begin{document}

% The following seven commands are intended for editorial usage and should be ignored by
% the author(s).
\Pagespan{789}{}% Document's page range. 
% If second parameter is left empty, the last page is computed automatically.
\Yearpublication{2006}%
\Yearsubmission{2005}%
\Month{11}%   
\Volume{999}%  
\Issue{88}% 
% \DOI{This.is/not.aDOI}% 

\title{RFI detection by automated feature extraction and statistical analysis}
   %\subtitle{subtitle?}

   \author{B. Winkel\inst{1}\fnmsep\inst{2}\fnmsep\thanks{Corresponding author:
  \email{bwinkel@astro.uni-bonn.de}}
          \and
          J. Kerp\inst{1}
          \and 
          S. Stanko\inst{1}
          }
\titlerunning{RFI detection by automated feature extraction and statistical analysis}
\authorrunning{B. Winkel,  J. Kerp \& S. Stanko}

   \institute{Argelander-Institut f\"ur Astronomie\thanks{Founded by merging of the Institut f\"ur Astrophysik und Extraterrestrische Forschung, the Sternwarte, and the Radioastronomisches Institut der Universit\"at Bonn.} (AIfA), Auf dem H\"{u}gel 71, 53121 Bonn\\
              \email{bwinkel@astro.uni-bonn.de};
              \email{jkerp@astro.uni-bonn.de}
              \and
              Max-Planck-Institut f\"{u}r Radioastronomie, Auf dem H\"{u}gel 69, 53121 Bonn, Germany
               }

\received{30 May 2005}
\accepted{11 Nov 2005}
\publonline{later}

% \abstract{}{}{}{}{}
% 5 {} token are mandatory

  \abstract
   {In this paper we present an interference detection toolbox consisting of a high dynamic range Digital Fast-Fourier-Transform spectrometer (DFFT, based on FPGA-technology) and data analysis software for automated radio frequency interference (RFI) detection. The DFFT spectrometer allows high speed data storage of spectra on time scales of less than a second. The high dynamic range of the device assures constant calibration even during  extremely powerful RFI events. The software uses an algorithm which performs a two-dimensional baseline fit in the time-frequency domain, searching automatically for RFI signals superposed on the spectral data. We demonstrate, that the software operates successfully on computer-generated RFI data as well as on real DFFT data recorded at the Effelsberg 100-m telescope. At 21-cm wavelength RFI signals can be identified down to the $4\sigma_\mathrm{rms}$ level. A statistical analysis of all RFI events detected in our observational data revealed that: (1) mean signal strength is comparable to the astronomical line emission of the Milky Way, (2) interferences are polarised, (3) electronic devices in the neighbourhood of the telescope contribute significantly to the RFI radiation. We also show that the radiometer equation is no longer fulfilled in presence of RFI signals.}
   {}

   \maketitle
%
%________________________________________________________________

\section{Introduction}\label{secintroduction}
The sky is not still any longer ``dark'' for radio astronomers. Tele\-com\-muni\-cation and radio broadcast are present at different power levels everywhere on Earth. The protected radio astronomical bands have offered the possibility to study the distribution of matter in space down to a certain sensitivity threshold. Today the sensitivity of the radio telescopes and their receiver systems has improved to detect continually fainter signals from distant sources. On the other hand, the protected bands for modern radio telescopes are becoming in parallel more polluted by signals from a variety of electronic devices which produce Radio Frequency Interference (RFI) across the entire electromagnetic frequency range.

In example, the leakage rate of a properly working microwave oven operating at 2.45\,GHz is typically about $1\,\mathrm{mW\,cm}^{-2}$ at a distance of 5\,cm. Considering the ITU-RA 769 recommendation for 21-cm line spectroscopy sets a threshold of $-196\,\mathrm{dB(W\,m^{-2})}$, all RFI-signals should be below this recommended power-level. Assuming that the first sub-harmonic contains only $1>p>10^{-3}$ of that flux density, the linear distance to detect this signal from a microwave oven is between 100\,km and 3\,km. Within this area several hundred houses are located in Central Europe. Moreover, we have also to consider the aging of the components of the electronic equipment, which might fulfil the strong regulations at the time of the first usage, but fails later on.
Moreover, an increasing fraction of technical equipment is needed to operate a state-of-the-art radio telescope. From a TFT-display to a GPS-clock all supplementary devices are located in the immediate vicinity of a radio telescope today and are at the same time potential sources of RFI emission.

Much of the contemporary activity in radio astronomy is focused on the exploration of the high redshift universe. These studies require observations at low frequencies with the highest possible sensitivities, far away from the protected radio bands, where most of the telecommunication, radio, and television broadcast stations are active. RFI mitigation thus is of at most importance to future radio telescopes like LOFAR (http://www.lofar.org) and the SKA (http://www.skatelescope.org).

In this paper we show that RFI mitigation is one of the major challenges in future radio astronomy. As a specific example, we studied the interference situation using the Effelsberg 100-m telescope at 21-cm wavelength. Our results indicate that RFI signals with highly variable intensities degrade the astronomical spectra across the whole protected band. Somewhat surprisingly, we find indications, that at least one major source of the RFI signals lies in the direction of the telescope control building. It is likely that the computer system and/or supplemental electronic devices produce harmonics across the 21-cm wavelength band.
One way to overcome such RFI degradation of the scientific data is to develop new data reduction methods.

Several methods have been developed today to detect or even mitigate RFI. In general, three different approaches can be identified. First is the ``classical'' method, a manual search for RFI signals in already recorded spectra.  Second, using sophisticated algorithms, it is possible to search for RFI signals in recorded data automatically (Bhat et al. 2005). Third, one can use real-time applications which have to be implemented into the signal chain of the telescope. Various approaches are under consideration, e.g. adaptive filters (Bradley \& Barnbaum 1996), post-correlators, or real-time higher order statistics (HOS; see Fridman 2001).

In this work we follow the second approach using snapshot data obtained with the high dynamic range DFFT spectrometer at the Effelsberg telescope (Stanko, Klein \& Kerp 2005). The method itself and its application to artificial spectra polluted by computer-generated RFI is presented in Sect.~\ref{secmethod}. In order to account for the quickly changing signal signature of most RFI events it is necessary to record the spectra on short time scales, at most a few seconds of integration. It became obvious during test observations, that the RFI signal variations at Effelsberg occur on the order of less than $100\,\mbox{ms}$. Such short integration times yield very high data rates. In practice, one must find a compromise between reasonable time-resolution for RFI detection and the amount of data. Moreover, the read-out time of the spectrometer is a technical constraint for RFI mitigation. The spectrometer and general experimental setup is discussed in Sect.~\ref{secsetup}.  In Sect.~\ref{secresults} we present the results of our RFI identification campaign at the Effelsberg 100-m telescope. Section~\ref{secdiscuss} compiles our summary and outlook for the future. To avoid confusion we introduce some naming conventions throughout the paper. A single spectrum is denoted as a \emph{scan}. A \emph{subset} denotes several spectra that refer to a specific position on the sky. All spectra from one observation (or all subsets) form a \emph{dataset}.

%__________________________________________________________________

\section{Algorithm}\label{secmethod}
\subsection{Feature detection and analysis}\label{subsecsurfacefit}
\begin{figure*}[!t]
%\centering
\includegraphics[width=\textwidth]{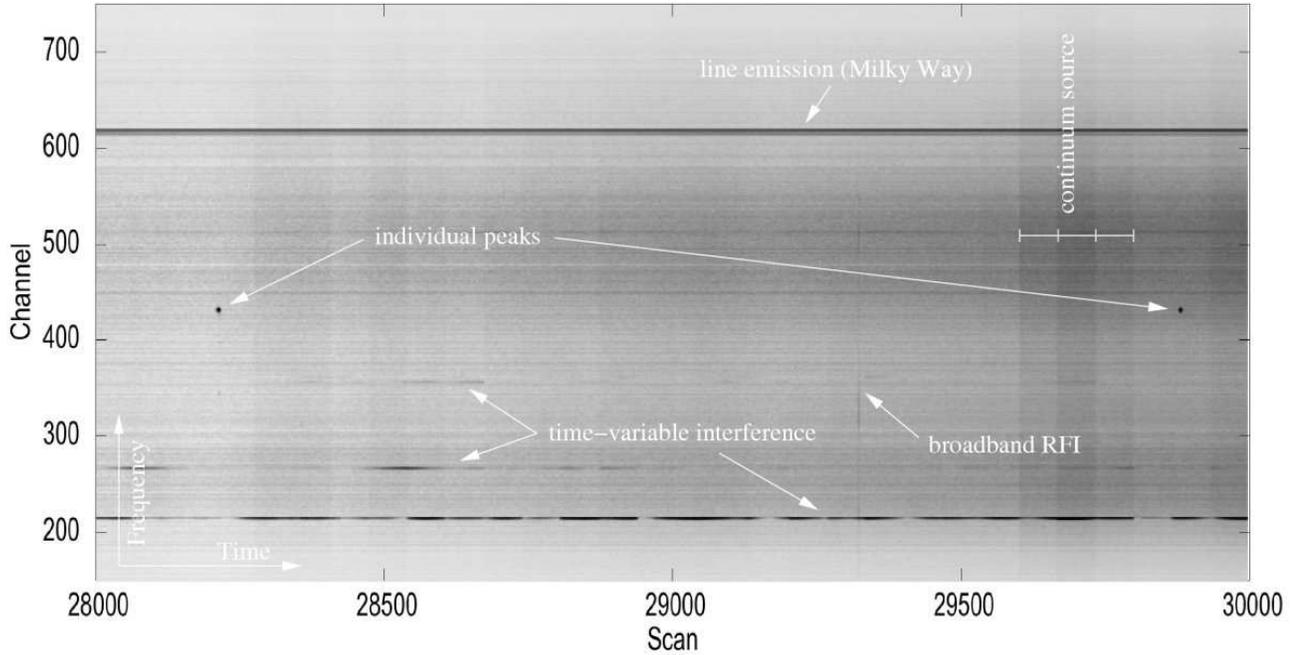}
\caption{Grey-plot visualising one of our datasets recorded with an FPGA-based DFFT spectrometer (Stanko et al. 2005) The data are composed of several subsets, each of about 30 single spectra containing data from a specific position on the sky. Three different types of RFI signals were detected. Note that due to its sensitivity the spectrometer reveals different continuum fluxes towards spatially separated positions.}
\label{figdatagreyplot}
\end{figure*}

Frequency, strength and location of RFI signals in astronomical spectra are assumed to follow a statistical distribution. The prime signature of RFI signals, in comparison to astronomical line emission, is the temporal variability. To make best use of this behaviour (Fisher 2002) suggested to use the so-called grey-plot which is a representation of the one dimensional astronomical spectra plotted as a function of time $t$. In Fig.~\ref{figdatagreyplot} a typical grey-plot of an astronomical \ion{H}{i} 21-cm line observation at the 100-m telescope is shown. The continuous and constant intensity line shows the \ion{H}{i} emission of the Milky Way. Throughout the whole grey-plot statistical distributed RFI peaks as well as long duration RFI signals can be identified. Based on our experience most types of RFI signals can easily be identified by eye in the time-frequency plane. This is because of the excellent pattern recognition ability of the human brain. 
However, hand\-ling several thousands of spectra is a task which demands pattern-recognition algorithms, which are also of great interest in many computer science problems such as classification of images, face recognition and many more applications. In principle, artificial neural networks (Duda, Hart \& Stork 2001) could be used for such an approach, but would need a huge amount of preprocessing and fine-tuning of the network which is outside the scope of this paper.

We implemented an automated feature detection supplemented with a follow-up statistical analysis which performs the separation between RFI signals and astronomical line emission. To optimise the RFI signal detection rate, the underlying envelope (henceforward denoted as baseline) should ideally be constant in both frequency and time domain.

This requirement is not fulfilled in real astronomical data. The shape of the baseline is determined by the bandpass of the receiver system. In the best case the shape of the bandpass is independent on the incident radiation power. As the flux of most cosmic sources can be considered as constant on timescale of minutes all changes of the baseline are due to drifts of the system, atmospheric effects, radiation from the ground, etc.  For subsets  with a duration of the order of one minute these changes can be assumed to be small, so that the functional dependence of the values in one spectral channel can be described by a low order polynomial. Usually, a baseline-fit is performed separately for each (long) integrated difference spectrum, which refers to a specific position in the sky. At this step in the reduction pipeline we are only interested in finding RFI signals, therefore we use (un-calibrated) total power spectra. A baseline-fit demands setting of so-called baseline windows, which define the spectral channels containing emission of cosmic sources. All spectral features of interest have to be covered within such a window, otherwise they would degrade the baseline-fit. Accordingly, RFI signals affect the baseline-fit strongly because in contrast to the emission of cosmic sources RFI signals are randomly distributed over time-frequency plane, so that in general no unique window can be defined. To overcome this limitation we developed a fitting procedure which automatically sets appropriate windows around spectral features, either of astronomical sources or of RFI signals. We separate the data into tiles -- parts of the frequency-time plane.  Each tile consists of data of 50~spectral channels (denoted as columns) of a single subset (in our case $\sim30$ individual spectra, denoting as rows). For simplicity reasons and to grant robustness versus fit-related computational uncertainties, we assume that each spectral channel in a tile has the same dependency of time. This means the baseline can be described by
\begin{equation}
f(\nu,t)=\sum_{i=0}^{\nu_\mathrm{order}} c_i\nu^i+\sum_{i=1}^{t_\mathrm{order}} d_it^i
\end{equation}
which represents a hyper-surface of $(\nu_\mathrm{order}+t_\mathrm{order}+1)$~degrees of freedom. This extends the typical 1-dimensional baseline fit procedure to a 2-dimensional fitting.

The algorithm automatically sets windows with a width of 5~pixels (5~spectral
channels times 5~successive scans) around all values above a trigger level of
$x_\mathrm{trig}\sigma_\mathrm{rms}$. As usual all data points within the windows are excluded from the baseline-fit. The fitting procedure and calculation of the residual are repeated until the fit has converged (reduced $\chi^2$ test). 
In practice we found that in general less than five iteration steps are sufficient. If very broad RFI signals are present the number of iterations is slightly higher because in each iteration the window may change its size only by a few pixels. A major issue on the robustness of this procedure is the first iteration. If one tile contains too many features, their initial impact on the fit could cause false results. To overcome this, we would need a first guess of window positions and sizes. This is done by applying horizontal/vertical matched filters $A_{\parallel,\perp}$ (edge-enhancement)
\begin{equation}
A_\parallel=\begin{pmatrix} 1 & 1 & 1& 1 & 1 \\ 0 &0& 0&0& 0 \\ -1&-1&-1&-1&-1 \end{pmatrix},~ A_\perp=\begin{pmatrix} 1 & 0 & -1 \\ 1 & 0 & -1 \\ 1 & 0 & -1 \\ 1 & 0 & -1 \\ 1 & 0 & -1 \end{pmatrix}
\end{equation}
We calculate the noise level (robust, by cutting off lower and upper 10\%~percentiles) and search for peaks ($5\sigma_\mathrm{rms}$ trigger level). This procedure sets initial constraints on the windows.

\begin{figure*}
\includegraphics[width=0.95\textwidth,clip]{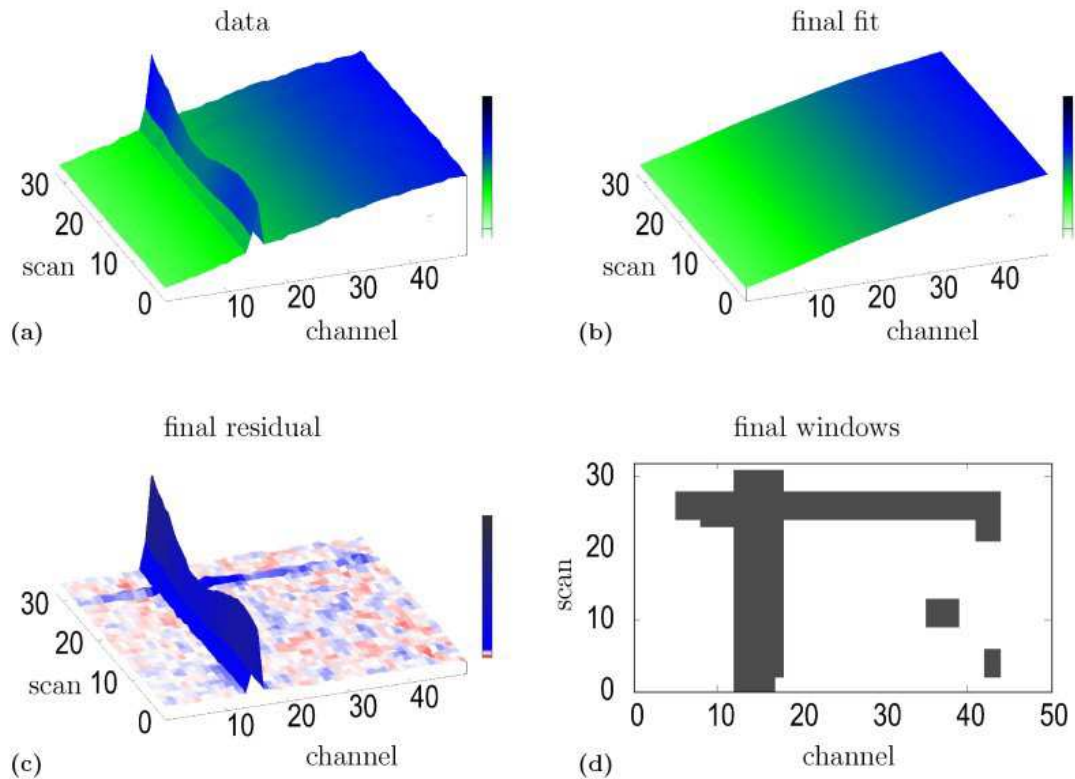}
\caption{\textbf{(a)} \ion{H}{i} 21-cm line data of the Effelsberg telescope containing a broadband interference superposing a narrow-width time-variable RFI signal. Shown is a part of a subset (tile) in time-frequency domain.  \textbf{(b)} Fit of the 2-dimensional baseline after five iteration steps to find the location of line emission and RFI signals in the data tile. \textbf{(c)} The difference between the data and the baseline in the time-frequency domain. The line emission and the noise are easy to identify. \textbf{(d)} Windows defined by the automatic window-fit algorithm.}
\label{figsampletile}
\end{figure*}

\begin{figure*}
%\centering
\includegraphics[width=0.9\textwidth,clip]{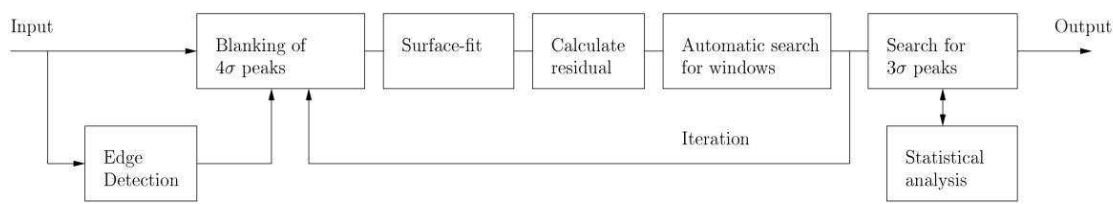}
\caption{Flow-chart of the detection algorithm. We use an edge-enhancement algorithm (utilising horizontal/vertical matched filters) to evaluate a first guess for features in time-frequency plane. In the successive iteration steps the fitted baseline is subtracted from the data. The result is inspected for signals in excess of a threshold $x_\mathrm{trig}\sigma_\mathrm{rms}$ which in return define new windows. The final residual contains only astronomical line emission and RFI signals. To distinguish between both a follow-up statistical analysis has to be performed; see Fig.~\ref{figstatanalysis}.}
\label{figflowchart}
\end{figure*}

Figure~\ref{figsampletile} shows a tile and the results of the automated feature detection procedure. In the final statistical analysis step all data points above a threshold level $x_\mathrm{thresh}\sigma_\mathrm{rms}$ which are enclosed within a window are taken into account. Note that the discrimination of trigger and threshold levels is in some cases helpful to avoid an unnecessarily high number of erroneous detections of noise peaks. The trigger level $x_\mathrm{trig}$ defines the intensity threshold to determine the extent of a window enclosing all pixels exceeding this threshold in the frequency--time domain. This general scheme is depicted in Fig.~\ref{figflowchart}.

\begin{figure}
%\centering
\includegraphics[width=0.45\textwidth]{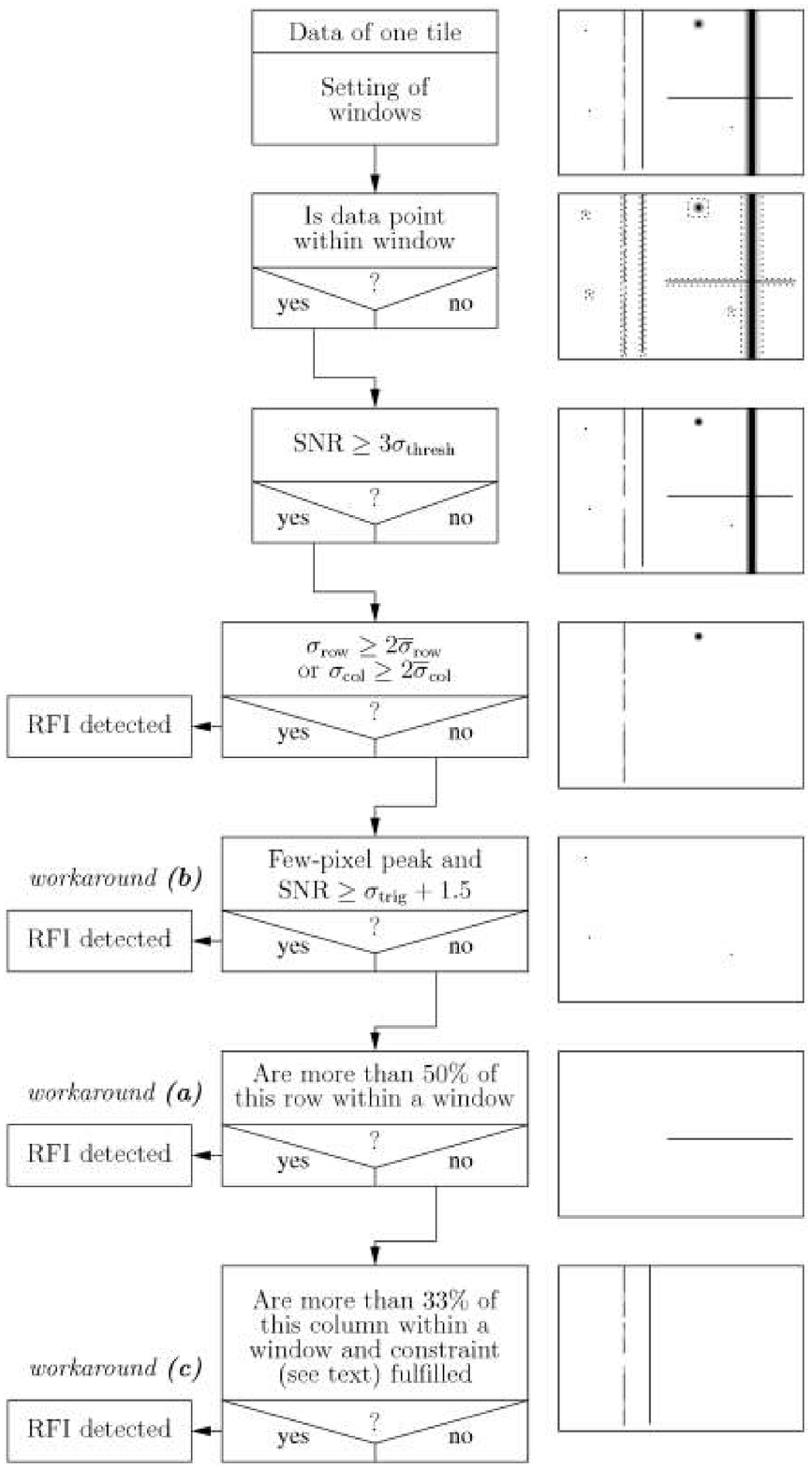}
\caption{The viewgraph shows the processing chain of the statistical analysis. After feature extraction all signals above a threshold level $x_\mathrm{thresh}\sigma_\mathrm{rms}$, which are localised in a window, are considered for follow-up statistical analysis. The main classifier uses variability of a signal, but to improve the detection quality we implemented additional morphology based classifiers.}
\label{figstatanalysis}
\end{figure}

The remaining task is to distinguish between RFI peaks and astronomical line emission within the windows. This is done by a statistical analysis; see Fig.~\ref{figstatanalysis} for a flowchart.The intensity of the astronomical line emission should approximately be constant within a subset. In the absence of RFI the standard deviations of the rows in a tile should have a similar value. The same holds true for the columns in the tile. As nearly all RFI signals are varying or modulated either in time or in frequency, an RFI event will change the statistical properties within the tile. Each data point within a window belongs to a specific row and column of the tile. If the standard deviation of this row or column is significantly (two times) higher than the median of the standard deviations of all rows/columns this is most likely due to RFI.

Such a statistical consideration does not work very well for relatively constant RFI signals, e.g. broadband interferences or impulsive but faint signals, as isolated weak peaks. Broadband interferences seem to be impulsive in time but enhance the continuum level across dozens or even hundreds of adjacent spectral channels. This easily triggers a window but is not sufficiently strong to enhance the standard deviation in a single tile significantly. To overcome this limitation, we implemented a workaround mechanism (referred to as \textbf{workaround (a)}), which is optimised to identify RFI signals polluting more than a half of the total spectral channels of a single row.  Isolated RFI peaks in the frequency-time domain are identified via \textbf{workaround (b)}: the algorithm searches for features in the tile which show up with an intensity in excess of $(x_\mathrm{thresh}+1.5)\sigma_\mathrm{rms}$ (to minimise the detection of noise peaks) which are narrower than the minimum expected line widths of astronomical sources and are short term events. At this stage the properties of the used backend have to be considered. Modern FPGA-based spectrometers will allow a high number of spectral channels and large bandwidth (Benz et al. 2005; Stanko et al. 2005). The combination of both determines the spectral resolution. In our case, we use 1024~spectral channels and 50\,MHz bandwidth which yields 50\,kHz equivalent to $\sim10\,\mathrm{km/s}$ at 1.4\,GHz (see Sect.~\ref{secsetup} for details).
The narrowest spectral line of interest should be sampled by at least three spectral channels. If a spectral feature shows up with a smaller line width as this lower limit, it is considered as an RFI-event. In the extreme case of 50\,kHz frequency resolution, the major fraction of the astronomical lines belonging to the cold neutral medium are not adequately sampled. Accordingly, we have to
account for these ``unresolved'' astronomical lines to differentiate them from
RFI-events.
Finally, a third workaround mechanism is implemented which allows to detect narrow-band RFI signals which have nearly constant intensities during the observation. Here, \textbf{workaround (c)} searches for channels which are associated with enhanced emission during at least 33\% of all scans. Because this also accounts for astronomical emission lines we use histograms of the pixels (within windows and above $x_\mathrm{thresh}\sigma_\mathrm{rms}$) vs. spectral channels. Therein we calculate equivalent widths of those candidates. If the signal has been identified as smaller than the expected minimal astronomical line width, we attribute this signal as an interference.

\subsection{Computer simulations}

\begin{figure}
%\centering
\includegraphics[width=0.45\textwidth]{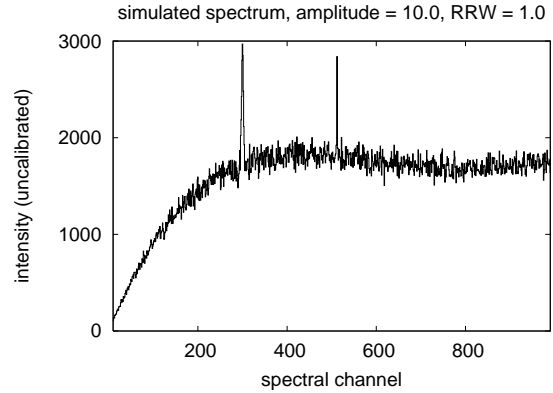}
\caption{Simulated spectrum containing noise, an ``astronomical'' line, and an interference peak. The result was multiplied with a polynomial of fifth degree to emulate the bandpass shape.}
\label{figbandpass}
\end{figure}

To determine the detection probabilities of the different RFI signal identification algorithms, we used simulated \ion{H}{i} spectra containing statistical noise, an irregularly shaped bandpass, and an ``astronomical'' line superposed by some randomly generated interferences (see Fig.~\ref{figbandpass}). The RFI signals were parametrised by (2-dimensional) sinc$^2$-functions (cutoff at first root) of adjustable width (interval between first roots, RRW) and scaled intensity (in units of SNR). This gave us the opportunity to test for a wide range of possible morphologies of the RFI signals. Note that due to sampling issues the constraints on minimum astronomical line widths were set to 1.5\,pixel although a narrow astronomical line at frequency resolution of 50\,kHz is also only represented by a single pixel.

\begin{figure*}
\includegraphics[width=0.9\textwidth]{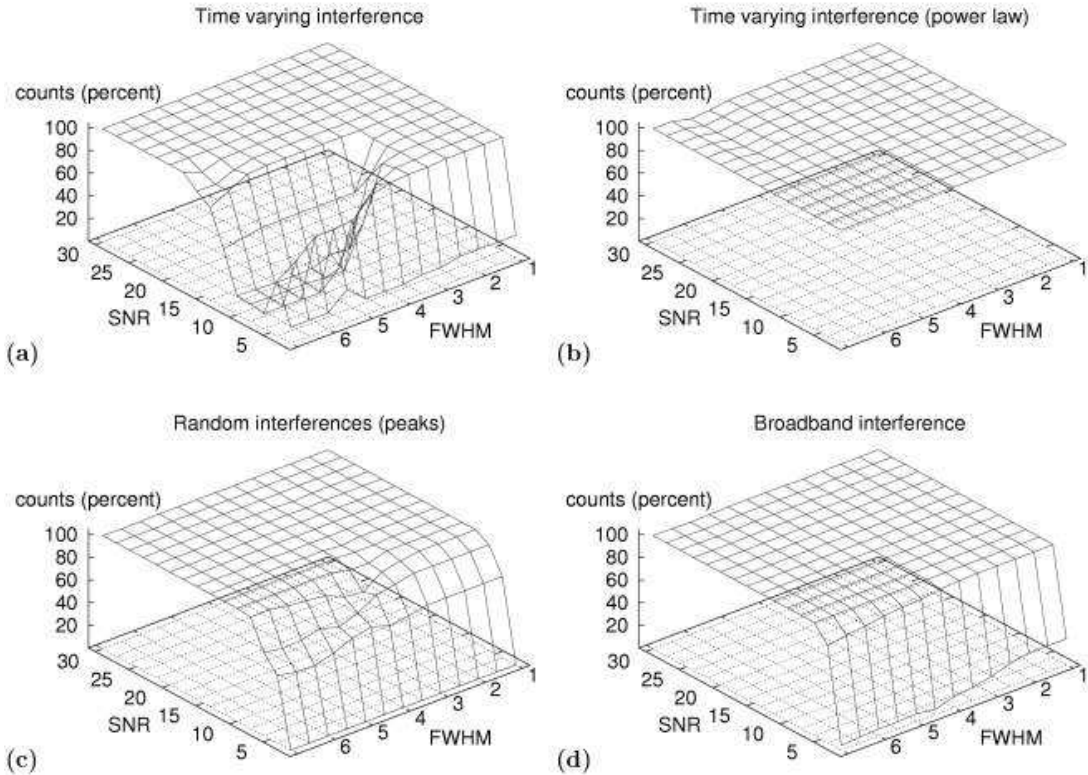}
\caption{Histograms showing the total detection rate of the proposed  algorithm. For each of the four simulated types of RFI -- \textbf{(a)} time-varying, \textbf{(b)} time-varying with underlying power law, \textbf{(c)} random peaks, and \textbf{(d)} broadband events -- the number of correctly found interference-containing data points was calculated as a function of signal-to-noise ratio (SNR) and root-to-root width (RRW, in pixels) of the $\mathrm{sinc}$-shaped specific events.}
\label{figsimulationsqualitytotal}
\end{figure*}

Four basic types of RFI signals were generated: 
(1)~We added a time-variable but frequency-stable RFI signal to each individual spectrum. This signal has a certain intensity modified by a sine modulation (with a period of 30~scans, which is slightly less than number of scans per subset) of 20\% of the mean interference intensity.
(2)~We simulated an RFI signal with variable intensities which are drawn randomly from a power-law distribution. In our observational data most of time-variable interferences show up with a probability distribution represented by a power-law with an exponent $\nu\sim-1.6$. If of narrow width, both time-variable types should be detected by workaround (c).
(3)~We simulated single peaks distributed randomly in both time and frequency. These kinds of RFI signals are detected by workaround (b) if they cover only a few pixels in the time-frequency domain. Otherwise their intensity or width is sufficiently high to enhance the variance statistically significant above the statistical noise level.
(4)~We generated broad signals which extend over a large number of channels with intensity modulations determined by a sine-wave. This type is likely detected by workaround (a).

Figure~\ref{figsimulationsqualitytotal} shows the detection rate of different RFI signal signatures as a function of both their amplitude (SNR) and width (RRW, in units of spectral channels or pixels, respectively). For the power law type the value of SNR denotes the base intensity of a constant signal onto which the power law values where added. Detection rate denotes the ratio of pixels suspected by the algorithm to contain RFI signals to the generated number of interference-polluted pixels. The corresponding trigger and threshold level (see Sect.~\ref{subsecsurfacefit}) were set to $x_\mathrm{trig}=x_\mathrm{thresh}=3$. In Fig.~\ref{figsimulationsqualitytotal}(a) one can clearly distinguish between the main (variance-driven) mechanism and workaround (c). Workaround (c) is responsible for the detection of interferences of lower SNR with narrow widths. These are already well-detected at a SNR of $3.5$. In general (for larger widths), non-power-law time variable interferences can only be successfully detected if the SNR is sufficiently high, because then the variance is also high enough. In our simulations a SNR of about 10 to 15\,$\sigma_\mathrm{rms}$ is needed. In  Fig.~\ref{figsimulationsqualitytotal}(b) 100\%~detection rate is reached because of the high variance of these kinds of signals. Moreover, the width of the RFI signal does not affect the detection rate as in the case of weaker sine-modulated signals. The bulk of the randomly distributed events (Fig.~\ref{figsimulationsqualitytotal}(c)) are well detected above a SNR-threshold of $6\sigma_\mathrm{rms}$. Again, there is no proportionality between the width of the RFI signal and its detection probability. Figure~\ref{figsimulationsqualitytotal}(d) shows the results for the broadband interference signals. Here the 100\% rate is reached at $4\sigma_\mathrm{rms}$. It is remarkable that even in the case of large widths (RRW) the algorithm is robust. At a width of 7~scans already 25\% of all values are affected! This robustness is only possible because of pre-filtering (edge-enhancement) the data as described above. 

\begin{figure*}
\includegraphics[width=0.9\textwidth,clip=]{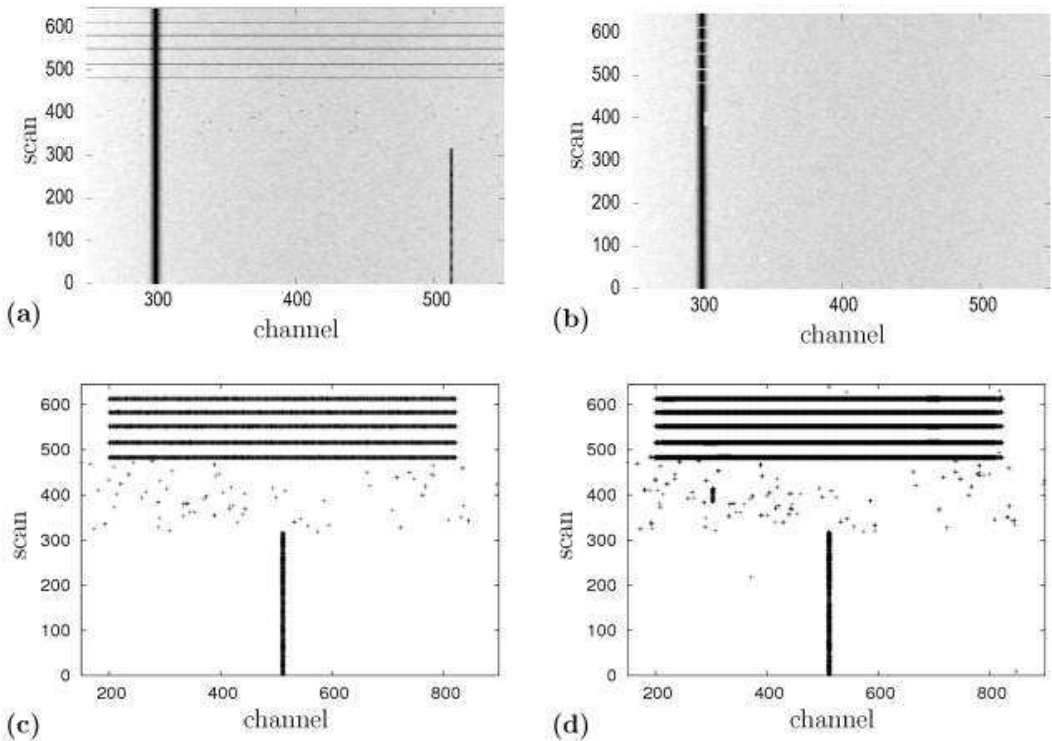}
\caption{Example, showing the results obtained from application of the proposed algorithm to simulated data ($\mathrm{SNR}=15$, $\mathrm{RRW}=3.0$). Four different types of interferences were added to the spectra. In panel \textbf{(a)} a grey-plot of the polluted data is shown with time-varying, random and broadband interferences. In the lower panel the detected peaks \textbf{(d)} vs. generated peaks \textbf{(c)} in time-frequency plane are shown. Panel \textbf{(b)} again shows a grey-plot but with cleaned data. Cleaning refers simply to substitute values in pixels, which are affected by an interference, with the value from the baseline fitting function.}
\label{figsimulations}
\end{figure*}

Figure~\ref{figsimulations} shows exemplary a generated  and cleaned dataset for an amplitude of 15 and a RRW of 3.0~channels. Cleaning denotes the substitution of the RFI enhanced spectral channels with values consistent with the neighbouring baseline function. We did not invest too much effort into a sophisticated cleaning procedure, because our primary intention was to detect interferences for flagging ``bad'' data. This has important consequences for automatic data reduction pipelines. Flagging of data leads to a loss data, but is by far better than the analysis of contaminated data (see also Fig.~\ref{figdatasensitivity}). The fast temporal sampling is necessary to identify the RFI affected pixels in the time-frequency domain and to differentiate those events from unresolved astronomical emission lines.

We want to discuss briefly the influence of integration time on the detection process. By calculating mean spectra the noise (RMS) is decreasing by a factor of $\sqrt{n}$ where $n$ is the number of spectra (assuming that all spectra have the same RMS). If a signal has constant strength in all of the spectra its SNR is increasing by the same factor. Thus, it is easier to detect. On the other hand, if a signal is short-lived that is it arises only in a single spectrum, its SNR will decrease by an overall factor of $\sqrt{n}$ when averaging. Most real events are mixtures of these extreme cases. But we can derive that short-lived events are harder to detect with increasing integration times while time-varying (long-lived) events are easier to detect. The latter, therefore, can be easily detected in mean spectra, but short-lived interferences are almost non-detectable unless they are very intense.

\subsection{Impact of RFI signals on sensitivity}
\begin{figure}
%\centering
\includegraphics[width=0.45\textwidth]{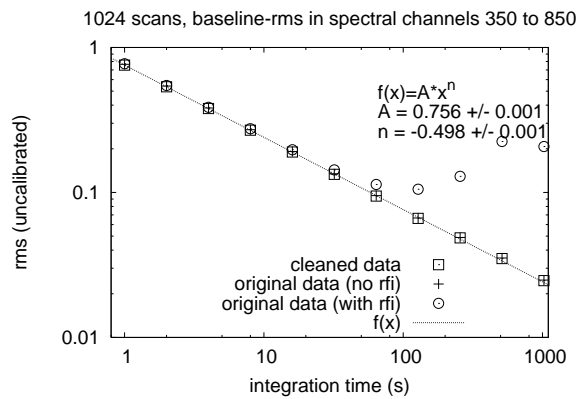}
\caption{Baseline RMS vs. integration time for $\mathrm{SNR}=10$ and $\mathrm{RRW}=1.0$. The theoretic noise level decreases with slope $\nu=-0.5$. The processed data yield this result, although the used cleaning algorithm is most simple. If no RFI-mitigation is performed after $\sim1\,\mathrm{min}$ integration time there is no further improvement on the sensitivity limit. Compared to typical integration times, of the order of several minutes, this becomes unacceptable.}
\label{figsensitivity}
\end{figure}

A further indicator for interference detection quality is given by the impact of the RFI signals and mitigation method on sensitivity. The noise level decreases with increasing integration time according to the radiometer equation
\begin{equation}
P(t)\sim \frac{1}{\sqrt{t\cdot\Delta f}}\sim t^{-0.5}
\end{equation}
with integration time $t$ and receiver bandwidth $\Delta f$. 

To evaluate the noise level as a function of integration time we generated 1024~spectra (scans). These contain the same four interference types as before. Figure~\ref{figsensitivity} shows RMS vs. integration time of simulated data with RFI, without RFI and cleaned. The term integration time has in principle no meaning for artificial spectra. We assume, that one scan refers to 1\,s as it is the case in our observations. The RMS is calculated by the baseline fit of a $4^\mathrm{th}$ order polynomial to the spectral channels 350 to 850. For simplicity reasons this range contains no simulated line emission, but many RFI signals. Next, we computed the mean of all RMS values as an estimator for the noise level in all scans. This is repeated for different integration times by calculation of the mean of each of two successive scans. After 10~steps only a single spectrum remains which is the mean of all 1024~spectra.

\begin{figure}
%\centering
\includegraphics[width=0.45\textwidth]{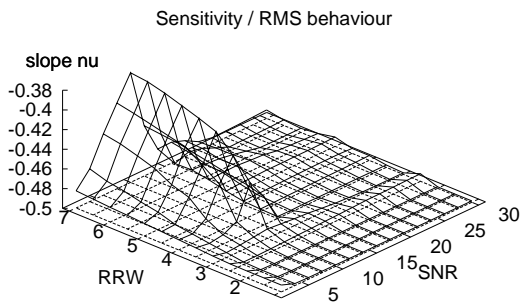}
\caption{Dependence of the slope $\nu$ (radiometer equation) on amplitude and width of simulated interferences. There is a small region in SNR-RRW plane where the theoretical sensitivity could not be reached. This is due to non-detection of broad time-varying (non-power-law) interferences; see Fig.~\ref{figsimulationsqualitytotal}(a). In all other cases the theoretical value of $-0.5$ could be reconstructed.}
\label{figslopeplot}
\end{figure}

One of the most important results is that without RFI mitigation the highest reachable sensitivity is far below that of clean data. On the other hand our proposed method is able to reconstruct the original RMS level extremely well; see Fig.~\ref{figslopeplot}. Here only a small region in the SNR-RRW plane is below the theoretical expected value.

%__________________________________________________________________

\section{Effelsberg 100--m telescope observations}\label{secsetup}
In Spring 2005 we used the Effelsberg 100-m telescope equipped with the 21-cm receiver combined with the new DFFT spectrometer backend (50\,MHz bandwidth, 1024 channels, two polarisation channels -- referred to as channel 1/3; see Stanko et al. 2005) to observe an area belonging to the halo of the Andromeda Galaxy. The aim of these observations was to search for very faint \ion{H}{i} 21-cm line emission associated with the giant stellar stream (Ibata et al. 2001). The area of interest was covered by four maps of $90\times30$ arc-minutes. Each map was measured 3 or 4\,times. This yields a total number of about 200\,000 scans which can hardly be analysed manually. Frequency and brightness temperature calibration were performed with the help of the standard calibration source S7 (Kalberla, Mebold \& Reif 1982). 

The field of interest is ideally suited to monitor in parallel the RFI signal intensities and frequencies during the observations. As mentioned above, it is necessary to find a compromise between integration time and storage time for each individual spectrum. We chose an integration time for each individual spectrum of 1\,s which yields an effective on-source time of 0.95\,s. We repeated the observation of each position 60\,times, gaining an effective observing time of about 2\,minutes by averaging both polarisation channels. The whole area of interest was mapped on a fully sampled grid. Each field of interest was observed at least three times. The observations were performed using the in-band frequency switching mode which was optimised to provide a sufficient number of spectral channels for the baseline evaluation and an appropriate frequency resolution of the data.

Direct sampling of the receiver signal, using a 100\,MHz A/D converter, is possible in the 0 to 50\,MHz band or in higher Nyquist zones of this band (under-sampling; see Stanko et al. 2005). In our observations we used the band between 150 and 200\,MHz. In principle we could use the 100 to 150\,MHz range as well, but there are some benefits from using the fourth Nyquist band. First, the receiver characteristic is slightly better. Second, we expect a higher interference rate in the third Nyquist band produced by FM broadcast at frequencies around 100\,MHz and in addition aircraft communications at somewhat higher frequencies. The intermediate frequency chain of the Effelsberg telescope is usually operating at 150\,MHz, so we had to shift the frequency of the local oscillator by 25\,MHz to 175\,MHz. 

\section{Results}\label{secresults}

\begin{figure}
\includegraphics[width=0.45\textwidth]{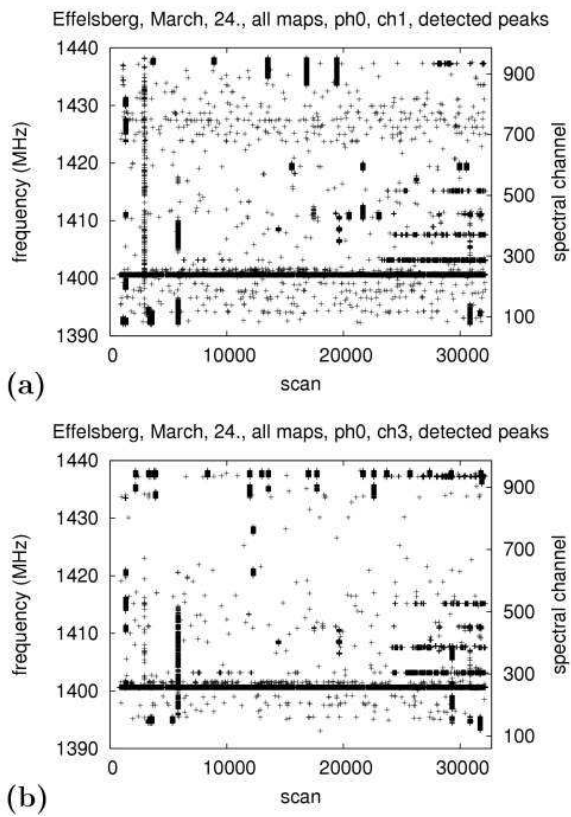}
\caption{Detected peaks in a part of the data (2005 March 24) for polarisation channels~1~\textbf{(a)} and~3~\textbf{(b)}. Although there seems to be a general similarity of both polarisations they differ locally from each other; see also Fig.~\ref{figchannel215}. }
\label{figdetected}
\end{figure}

We analysed the Effelsberg 21-cm line data using $x_\mathrm{trigg}=3$ and $x_\mathrm{thresh}=3$ for the detection thresholds. Note that we searched for interferences in both frequency switching phases (ph0 and ph1) separately. Figure~\ref{figdetected} shows exemplary the time-frequency plane of a typical observation for polarisation channels 1~(a) and 3~(b). The crosses mark the locations of detected signals identified as RFI. Most of them are produced by real interferences. If a significant number of detections would be caused accidentally by noise peaks we would expect a uniform distribution of such signals in time-frequency plane, which is not the case. A single RFI signal in general affects multiple pixels in time-frequency domain because frequently the neighbouring spectral channels and scans are also contaminated. The software stores the localisation of all RFI signals in time-frequency domain, their observational parameters, and the SNR of each peak in the corresponding tile. This information can be used to clean the observational data and for statistical evaluation.

\subsection{Polarisation and sensitivity}
\begin{figure}
\includegraphics[width=0.45\textwidth]{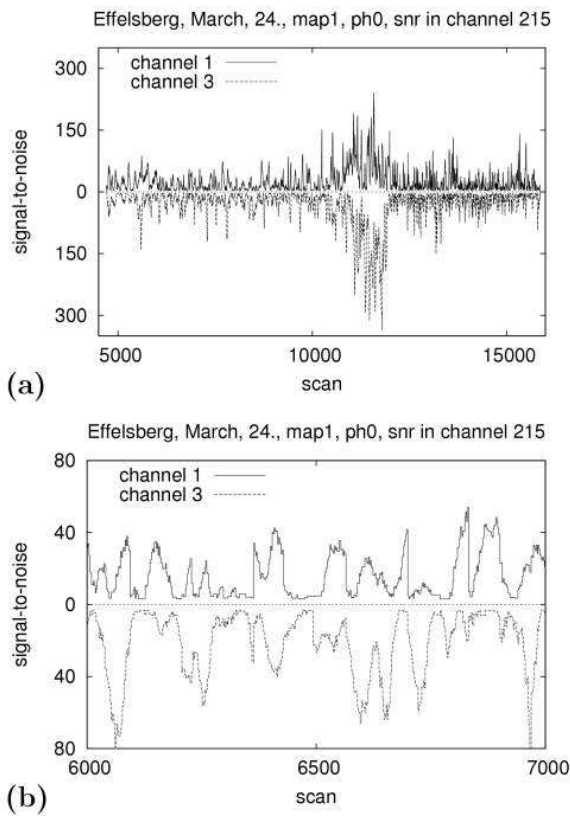}
 \caption{\textbf{(a)} SNR for the spectral channel~215 of a part of the data (2005 March 24). Panel \textbf{(b)} shows a zoom-in. There is a similar global behaviour but locally the SNR of both polarisations differ, meaning that interferences are linearly polarised but amplitude and/or phase relation changes randomly.}
\label{figchannel215}
\end{figure}

One of the most interesting aspects of RFI signals is how they enter the receiving system. On one hand, if they enter the system via the antenna, their polarisation properties might vary statistically. On the other hand, if they enter the system directly via the intermediate frequency chain, the polarisation properties are expected to be stable during most of the observing time, as then both polarisation channels are differently affected. We find, that the absolute majority of all RFI signals are polarised. To illustrate this finding, we plot in Fig.~\ref{figchannel215}(a) the signal of a single spectral channel as a function of time. The solid line represents the left-hand and the dotted line marks the right-hand polarised signal. The intensity variations of both polarisation channels are comparable. However, if we zoom into the temporal variation of the signals (Fig.~\ref{figchannel215}(b)) we can identify major differences between both. Apparently, both polarisations have a different evolution. Most of the RFI signals are linearly polarised (or got polarised by interaction with the telescope dish) and couple into the receiving system, equipped with a circular feed, in a statistical manner. The same is true for the other spectral channels which is suggested by Fig.~\ref{figdetected}. Globally both polarisations show a similar event density but differ locally from each other.

\begin{figure}
\includegraphics[width=0.45\textwidth]{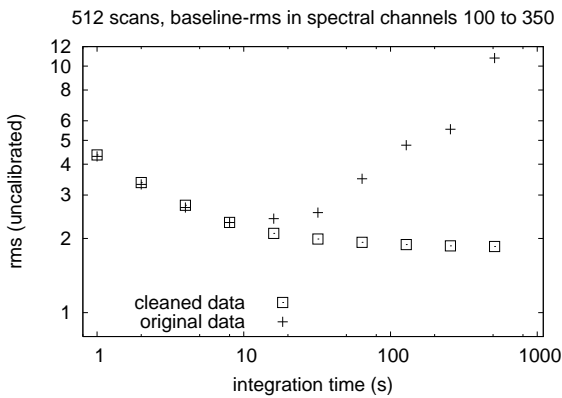}
 \caption{In the presence of an RFI signal (in this case time-varying and narrow-band) the reachable sensitivity is limited. Because of the varying RFI signal strength, the RMS of the uncorrected data grows even with integration time.}
\label{figdatasensitivity}
\end{figure}

In Fig.~\ref{figdatasensitivity} we show exemplarily the sensitivity as a function of integration time in the presence of the specific time-varying RFI event in spectral channel~215. Figure~\ref{figdatasensitivity} was computed by subtracting the baseline in the spectral interval from channel~100~to~350 and calculating the mean RMS of the residual for different integration times. Again, the RMS in the presence of an RFI signal shows strong deviation from the radiometer equation, as our simulations suggested.

\subsection{Frequency of occurrence and mean signal strength}
\begin{figure*}
\includegraphics[width=0.95\textwidth]{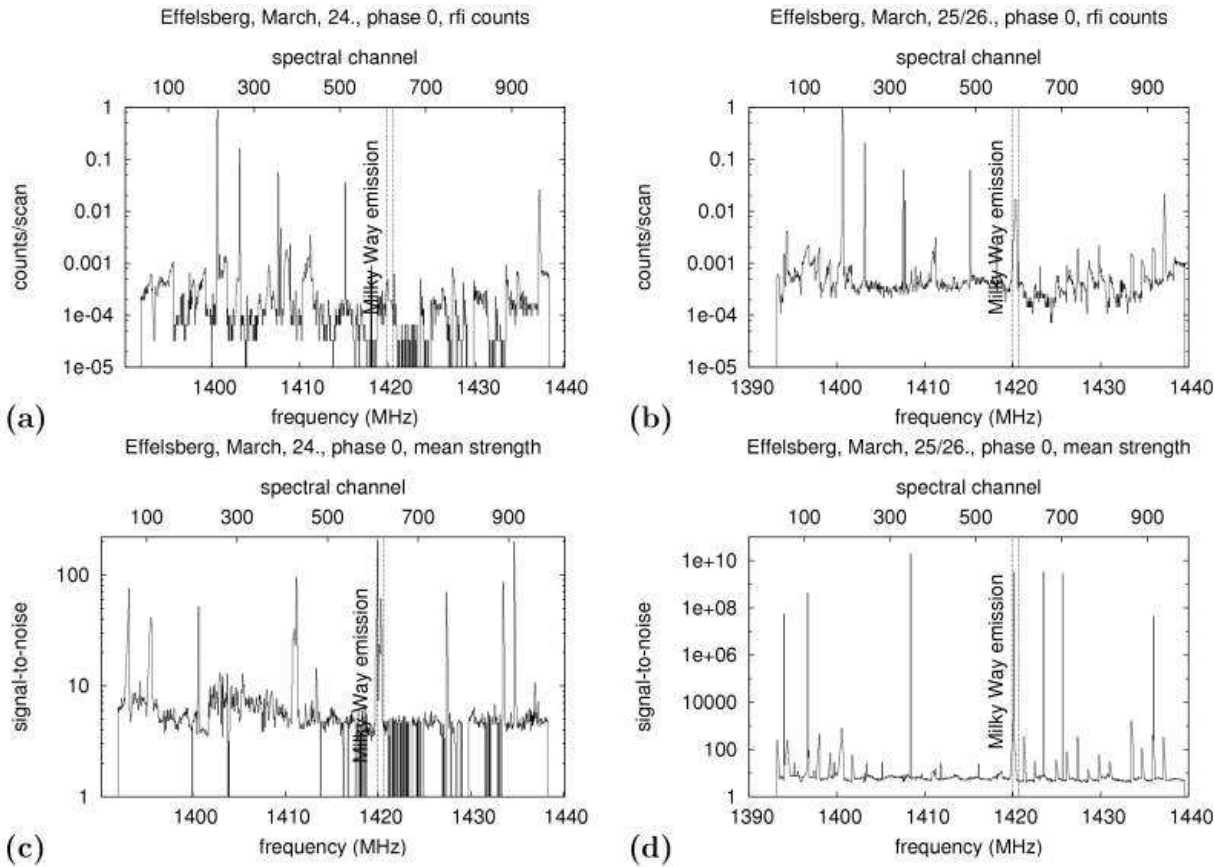}
\caption{Frequency of occurrence and mean SNR of all interferences in both polarisation channels in phase~0. The upper part shows the absolute counting rates normalised to counts per scan and spectral channel. The lower part shows the mean SNR of all counted peaks per spectral channel. Only in the second half of the observation extremely intense impulse-like events occurred. These have SNRs being several orders of magnitude higher than in all other events. Figure~\ref{figmegaevent} shows the \ion{H}{i} spectrum of such an event.}
\label{figstatmapcounts}
\end{figure*}

Figure~\ref{figstatmapcounts} shows the frequency  of RFI signals as a function of spectral channel. The count rates were normalised with respect to the total number of scans. The values represent accordingly the relative occurrence of an RFI signal per scan. The lower panel of Fig.~\ref{figstatmapcounts} shows the mean interference signal strength per spectral channel normalised to the number of counts per spectral channel. Some RFI signals (i.e. at 1400.6\,MHz) are detected in almost every individual scan. The mean strength of the signal at 1400.6\,MHz is about 100$\sigma_\mathrm{rms}$ ($\sim9$\,K) which makes it almost as intense as the line emission of the Milky Way ($\sim11$\,K). The line emission of the Milky Way is located in the frequency range between 1420 and 1422\,MHz. It is remarkable that our algorithm apparently distinguishes well between line emission and interferences, as only very few real astronomical lines were accidentally counted. 

\begin{figure}
\includegraphics[width=.45\textwidth]{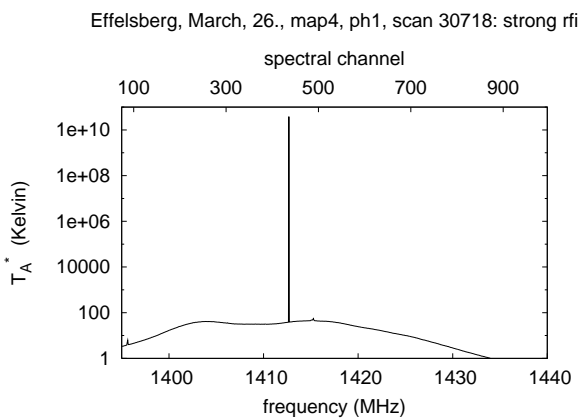}
 \caption{Spectrum containing a very short and narrow but extremely intense interference. The backend was not overloaded due to the high dynamic range (14\,bit) of the spectrometer. The spectrum was calibrated in terms of brightness temperature $T_\mathrm{A}^\ast$. The signal reaches intensities in the order of $10^{10}$\,K.}
\label{figmegaevent}
\end{figure}

\begin{figure*}
\includegraphics[width=.95\textwidth]{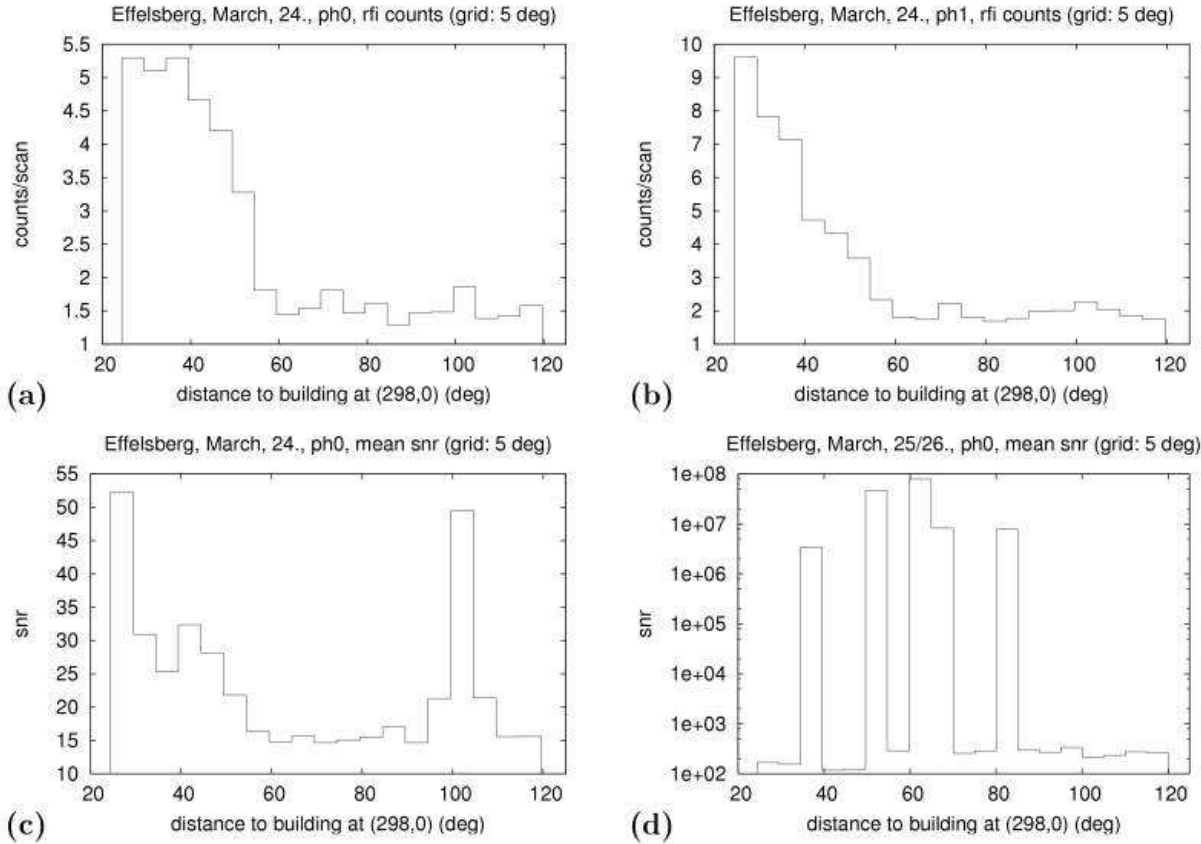}
\caption{RFI counts as a function of angular distance to the observatory building at Effelsberg. Panel \textbf{(a)} and \textbf{(b)} show the count rate which significantly increases for short distances. A local maximum at $100\degr\ldots105\degr$ could originate from spill-over. The spill-over angle of the 100-m-telescope lies in the range of about  $90\degr\ldots95\degr$. Panel \textbf{(c)} and \textbf{(d)} show the mean SNR of the interferences vs. angular distance. A similar behaviour is found. The distribution of the events with extreme intensities does not show a determinable pattern, as can be seen in panel \textbf{(d)}.}
\label{figdistance}
\end{figure*}

Because of technical reasons we slightly changed the switching frequency in the second half (2001 March 25/26) of the measurement. On these days in addition to the ``normal'' RFI signals also extreme events have been observed (Fig.~\ref{figmegaevent}). The brightness temperatures of these extreme events are up to $10^{10}$\,K. Because of the unusually large dynamic range (14\,Bit) of the DFFT spectrometer it was feasible to observe without the necessity of re-calibrating the receiving system when such an event occurred. The presence of these events is probably unrelated to the different switching frequency but is due to a different observation date.

\subsection{Dependence on observational parameters}
Using the number counts of the detected RFI signals it is possible to constrain the direction of their origin. We plotted in Fig.~\ref{figdistance}(a) the frequency of the RFI signals as a function of the angular distance from the building. 

A local maximum of RFI signals is reached when the telescope points closest towards the control building. At an angular distance of about 100\degr a secondary local maximum can be identified. This second maximum becomes more pronounce if we convert the count rate of RFI signals into SNR; Fig.~\ref{figdistance}(b). We attribute this second maximum to the so-called spill-over ring of the telescope receiver system. The 21-cm receiver is mounted within the primary focus. It is known, that some radiation enters into the 21-cm receiver directly from the ground. At an angular distance of 100\degr the RFI signals produced within the control building enters directly into the receiver system. This finding shows in addition that most of the RFI signals enter the telescope-receiver system via the telescope and not directly via the intermediate frequency chain. The latter is also approved by the fact that we have not found interferences which do not follow frequency switching, as it would be the case if some RFI signals entered after mixing to IF.

The extremely intense events, which only occur in the second half of the measurements, show no recognisable angular dependence, but seem to be randomly distributed. Beside, such events are very rare.

\subsection{Distribution}
\begin{figure}
\includegraphics[width=0.45\textwidth]{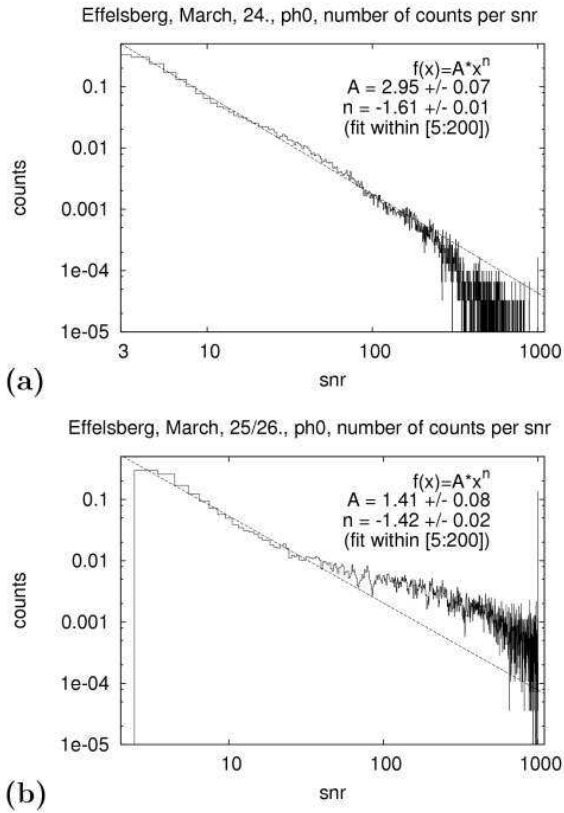}
\caption{Number of peaks with a certain SNR. The number distribution can be described by a power law $f(\sigma)=A\sigma^\nu$ with $\nu\sim-1.5$ over a wide range of SNRs.}
\label{figpowerspectrum}
\end{figure}

Finally, we calculated the distribution of RFI signals as a function of SNR. In Fig.~\ref{figpowerspectrum} we plotted in logarithmic scale the histogram of the SNR of the detected peaks. Over a wide range of $\sim4\ldots200\sigma$ SNR the distribution can be described by a simple power--law $A\sigma^\nu$ with $\nu\sim-1.6$. The nature of the process causing this power law is unknown. It is mainly associated with the time-varying interferences, as these represent more than 95\% of all events. From test observations at 18-cm we know another type of interference (probably originating from the GLONASS satellite) which yields a Gaussian-like distribution of signal strengths. It is likely that each different type of RFI source has a typical distribution, which in turn could help to distinguish them.

%_____________________________________________________________

\section{Summary}\label{secdiscuss}
We developed a new algorithm for an automated RFI signal detection in existing 21-cm line data. These had to be recorded using short integration times of less than 1\,s. Most of the existing spectrometer backends do not allow such a fast storage of spectra. The new DFFT spectrometer which was developed at the University of Bonn by (Stanko et al. 2005) fulfils this need. Furthermore it provides a large dynamic range of 14\,bit so that it is best qualified to do RFI analysis. But in principle the proposed algorithm could easily be adapted to other spectrometers, if they were capable of delivering short-integrated spectra. The time variability of most interferences demands analysis on short time scales. This in return means a huge amount of data to handle which can hardly be done manually. This becomes especially true in case of surveys where several millions of spectra have to be analysed if RFI mitigation on such time scales is aimed for.

The application of the algorithm to simulated data revealed its capability to find very different types of interferences with a high detection rate. It could be shown that even at low signal intensities of about $4\sigma_\mathrm{rms}$ a large fraction of all events was found. Probably most remarkable is that without RFI mitigation whether automatically or not, the radiometer equation is no longer fulfilled. The proposed algorithm could greatly improve this issue. It is likely that we will find other morphological types of RFI (radar chirps etc.) in upcoming observations. If these turned out to be not well-detectable by our current software, further workaround procedures would be easy to implement due to its modular design.

We developed a tool for doing statistical analysis of RFI events, which turned out to be very useful for working out properties of interferences and helpful in finding possible interference sources. Analysis of the data recorded at the 100-m Effelsberg telescope yielded some interesting results. Most of the RFI signals are linearly polarised. We could show that the majority of RFI signals enters the telescope via the telescope antenna and not via the intermediate frequency chain. Also, most of the detected RFI signals are produced by the electronic devices within control building of the 100-m telescope. The most extreme RFI signals appear to enter the telescope from randomly distributed directions. 

Note, that the large counting rate from the observatory in Effelsberg is not due to bad technical conditions. Radiation at such low power levels is quite common. Soon most of electronic equipment at Effelsberg will be placed in a Faraday cage. We are curious how this will improve the situation.

%_____________________________________________________________

\acknowledgements
We would like to thank Tobias Westmeier for very helpful discussion on the manuscript. Based on observations with the 100-m telescope of the MPIfR (Max-Planck-Institut f\"{u}r Radioastronomie) at Effelsberg.

\end{document}